\documentclass[preprint,12pt,3p]{elsarticle}
\makeatletter
\def\ps@pprintTitle{%
 \let\@oddhead\@empty
 \let\@evenhead\@empty
 \def\@oddfoot{}%
 \let\@evenfoot\@oddfoot}
\makeatother

\usepackage{amssymb}
\usepackage{amsmath}
\usepackage[]{hyperref}

\begin{document}

\begin{frontmatter}

\title{A minimalist model for co-evolving supply and drainage networks} 

\address[label1]{Department of Civil and Environmental Engineering, Princeton University, USA}
\address[label2]{ Princeton Environmental Institute, Princeton University, Princeton, USA}
\address[label3]{Princeton Institute for International and Regional Studies, Princeton University, USA}
\address[label4]{Department of Mathematics, University of Bergen, Bergen, Norway}

\cortext[cor1]{I am corresponding author}

\author[label1]{Shashank Kumar Anand}% \fnref{label3}}
\ead{skanand@princeton.edu}
\author[label2,label3]{Milad Hooshyar}
\ead{hooshyar@princeton.edu}
\author[label4]{Jan Martin Nordbotten}
\ead{jan.nordbotten@math.uib.no}
\author[label1,label2]{Amilcare Porporato \corref{cor1}}
\ead{aporpora@princeton.edu}

\begin{abstract}
Numerous complex systems, both natural and artificial, are characterized by the presence of intertwined supply and/or drainage networks. Here we present a minimalist model of such co-evolving networks in a spatially continuous domain, where the obtained networks can be interpreted as a part of either the counter-flowing drainage or co-flowing supply and drainage mechanisms. The model consists of three coupled, nonlinear partial differential equations that describe spatial density patterns of input and output materials by modifying a mediating scalar field, on which supply and drainage networks are carved. In the 2-dimensional case, the scalar field can be viewed as the elevation of a hypothetical landscape, of which supply and drainage networks are ridges and valleys, respectively. In the 3-dimensional case, the scalar field serves as the chemical signal strength, in which vascularization of the supply and drainage networks occurs above a critical `erosion' strength. The steady-state solutions are presented as a function of non-dimensional channelization indices for both materials. The spatial patterns of the emerging networks are classified within the branched and congested extreme regimes, within which the resulting networks are characterized based on the absolute as well as the relative values of two non-dimensional indices.
\end{abstract}


\end{frontmatter}

% \linenumbers

%% main text

\section{Introduction}
Many natural and man-made systems consist of materials being conveyed in and/or out of the domain through preferred routes, which result in the evolution of supply and/or drainage networks. In some biological systems, motile cells regulate their movement based on the affinity or aversion to specific environmental factors (temperature, chemical/biological signal) \cite{hedgecock1975normal,manoussaki1996mechanical,chaplain2000mathematical, hillen2009user}. Two co-existing materials, moving up and down a signal gradient, drive the formation of the competing networks. In other systems, the material is supplied throughout a domain and gets collected once it has been utilized (and often also transformed), resulting in the formation of co-flowing supply and drainage networks. Examples include the cardiovascular network of blood and nutrients in animals, the supply-chain network of a commodity from the manufacturer to the customer and the related disposal, the aqueduct, and waste-flow network in urban water systems \cite{rinaldo1993self,sun1995minimum,dandy1996improved,abdinnour1999network, suweis2011structure,mahlke2007simulated,bonetti2020channelization}. In all these systems, the co-existing networks must evolve or be designed in a way that is coordinated, depending on different constraints, such as the configuration of the distribution region, the cost and modes of transportation for supply and drainage material, etc. 

A great deal of research has explored the quantitative laws that explain the structure of networks in different disciplines, but this has been typically done considering either the supply or the drainage network separately \cite{banavar1999size, hooshyar2017hydrologic, minoux1989networks, banavar2000topology,ronellenfitsch2016global}. In many cases, the general framework for studying such systems has been a static cost optimization problem typical of optimal transport theory \cite{dorogovtsev2002evolution, bohn2007structure, danila2006optimal, durand2006architecture}. As a result, the topology of the underlying supply or drainage network depends on the definition of the cost, including minimum energy dissipation, geometrical constraints, etc. \cite{chen1997simple,rodriguez2001fractal,murray1926physiological,rachev1985monge}. Recently, there have been efforts to provide the interpretation of this static principle as the result of a dynamic evolution based on partial differential equation (PDE) \cite{evans1997partial,facca2018towards,cardin2019optimal,hooshyar2020variational}.

Less efforts have been devoted to analyze the co-evolution of supply and/or drainage transport systems within a continuous domain, which is complicated by the presence of common and individual factors that affect transportation for both materials, including shape and size of the region, extra scalar/vector field, production/consumption rate, velocity, etc. As a step in this direction, this study aims at formulating and analyzing a minimalist model that captures the essential interactions between two materials being conveyed in a continuous domain, where the system can be interpreted either as a counter-flowing drainage system or a co-flowing supply and drainage system. The model can be generalized to incorporate multi-species interplay; however, we keep the discussion up to two-species interactions.

The conceptual framework developed here stems from observing the complex ridges and valleys patterns in topographic landscapes, and the related work in the fields of image processing, geomorphology, and hydrology to formalize the duality between the interlocking network of ridges and valleys \cite{koenderink1993local, werner1991several, dawes1994significance}. For mathematical formulization, we draw inspiration from landscape evolution models (LEMs) which have been successful in describing the formation of river and stream networks \cite{fowler2011mathematical,istanbulluoglu2005vegetation, bonetti2020channelization, perron2012root}. Generalizing these models, we develop a simple system consisting of three nonlinear coupled PDEs with the essential parameterization. We introduce a scalar field in a continuous domain that mediates two competing mechanisms of two counter-flowing drainage or co-flowing supply and drainage. We show the influence of rules of production and/or consumption as well as the boundary conditions on the obtained steady-state network patterns. The two channelization indices for both materials are obtained by the non-dimensionalization of the coupled PDEs, which allow us to define the role of various common and individual factors on the extent and patterns of the formed networks.

The paper is structured as follows. In Section \ref{S2}, we first present the conceptual framework for both viewpoints of the model. We construct the 3-field mathematical model and define non-dimensional indices to describe the relative importance of various factors that alters the characteristics of the coupled networks. We also show that for unity value of exponents, the proposed model can be re-written as a 2-field model. The steady-state closed-form solutions for non-channelized flows in 2 and 3 dimensions are derived in Section \ref{S3}. In Section \ref{S4}, the numerical simulation results for the 2-dimensional and 3-dimensional cases are presented and the spatial patterns are analyzed for different levels of complexity and branching. Conclusions and future research directions are discussed in Section \ref{S5}. In \ref{2field}, we discuss the 2-field equivalent formulation for the proposed PDE model and the complexity in the boundary conditions that emerges from this model reduction.

\section{\label{S2} Mathematical model}

\subsection{Conceptual model inspired by the ridge and valley duality}
\label{S21}

\begin{figure}[!hbt]
\centering
\includegraphics[width=0.9\linewidth]{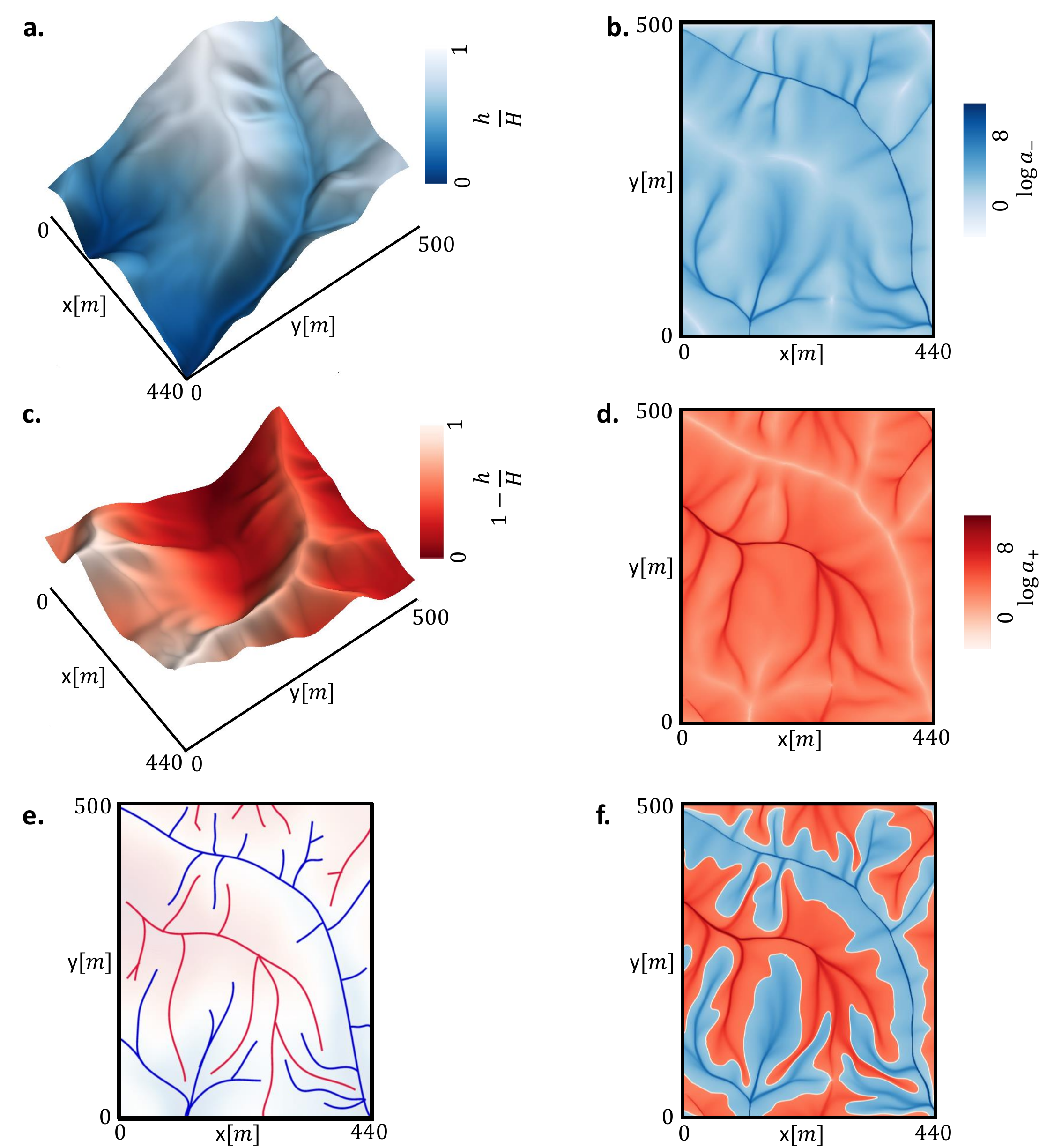}
\caption{Conceptualization of supply and drainage networks using the dual ridge and valley networks in a topographic landscape. (a): 3-dimensional surface ($h$) for the selected topography. (b): Drainage network for $a_-$ following the negative gradient of $h$. (c): The inverted 3D surface for the original topography, where $H$ is the maximum elevation in the domain. (d): Drainage network for $a_+$ following the negative gradient of the inverted $h$. (e) Interlocked planar ridge and valley networks with prominent ridge-lines (red) and valley-lines (blue). (f) The white curve represents the interface $a_+ = a_-$ and separates the red-colored region representing high accumulation of $a_+$ ($a_+>a_-$) from the blue-colored region showing aggregation of $a_-$ ($a_->a_+$). The selected topography is from the Calhoun Critical Zone landscape in South Carolina (obtained from the \href{https://opentopography.org/}{the OpenTopography facility).}}
\label{fig:one}
\end{figure}

We begin by considering the geometry of a topographic field (Figure \ref{fig:one}a), visualized as a scalar field ($h$) in 3-dimensional Euclidean space, where the vertical direction points in the direction of gravity. Avoiding maxima (summits) and minima (pits), the curves of particular significance here are the ridges and valleys, which provide a skeleton for the structure of the drainage network \cite{koenderink1998structure, bonetti2018theory}. With the assumption of negligible inertial effects, a fluid present in the domain flows under gravity over the scalar field along the direction of the steepest descent, resulting in the distribution of material density, say $a_-$, as shown in Figure \ref{fig:one}b,
highlighting the drainage network for the topography. This density, $a_-$, is drained by the stream network and flows out of the system at the boundary.

Inverting artificially the initial topography, as shown in Figure \ref{fig:one}c, the duality between ridges and valleys is apparent, as ridges become valleys and valleys become ridges. The interlocked network of ridge-lines and valley-lines extracted from the original topography is shown in Figure \ref{fig:one}e. Based on this duality, and similarly to the density of $a_-$ for the drained material, one can imagine another flow with density $a_+$ in this inverted topography, which is produced within the domain and gets drained by the ridge network. The field of material density ($a_+$), marking the drainage network for the flipped topography, is shown in Figure \ref{fig:one}d, where the main courses of flow follow the ridge-lines of the original topography. Therefore, the flow of $a_+/a_-$ moving up/down the slope of the topographic field to be drained by the ridge/valley network becomes the counter-flow problem (we will refer to this as Problem I).

Reversing the flow direction of the density $a_+$ in above scenario, the problem can be formulated as a co-flowing supply-drainage problem (Problem II), where $a_+$ represents the density of the supplied material that flows down the slope similar to the drained material ($a_-$). Instead of having a distributed source through the domain and exiting from the boundary through the ridge network, $a_+$ enters from the boundary where the ridge forms peak, and flows along the ridge network following the topographic steepest descent. Figure \ref{fig:one}f displays the accumulation of $a_+$ along ridges (red-colored region) and accumulation of $a_-$ along the valleys (blue-colored region), with the white curve subdividing regions dominated by either material. One can envision the supplied material (density, $a_+$) enter the area at the boundary concentrated at the ridges of the scalar field ($h$), flowing and getting distributed over the hillslopes as it gets exhausted. In turn, the consumption of the supplied material produces the drained material (density, $a_-$), which moves under the scalar field potential, and gets discharged out of the domain preferentially via the valleys.

\subsection{Governing equations}
\label{S22}

The 2-dimensional illustration presented above can be formalized and extended to an $n$-dimensional space ($\mathbf{R}^n$), considering a scalar field $h:\mathbf{R}^n \rightarrow \mathbf{R}$, defined inside a domain $\Omega$ along with two scalar fields $a_+$ and $a_-$ playing the role of the densities of materials. 

For the counter-flow drainage problem (Problem I), the continuity equation for the two materials ($a_+$ and $a_-$), that are produced at a unitary rate and flow with opposite velocity $\mathbf{v_+}$ and $\mathbf{v_-}$, respectively, can be written under the assumption of quasi steady-state as
\begin{equation}
\nabla \cdot \left(a_\pm \mathbf{v_\pm}\right) = 1, \label{a_pm_eq_dd}
\end{equation}

For simplicity, we assume that the velocity fields, $\mathbf{v_+}$ and $\mathbf{v_-}$, follow the positive and negative gradient of $h$, respectively, with unit speed as
\begin{equation}
\mathbf{v_{\pm}} = \pm\frac{\nabla h}{|\nabla h|} \label{v_pm_eq_dd}.
\end{equation}

The scalar field $h$ is assumed to co-evolve with the fields of both materials. Specifically, the temporal evolution of $h$ consists of a diffusion term and nonlinear, nonlocal sink and source terms due to the feedback from both materials as
\begin{equation}
\label{h_eq}
\frac{\partial h}{\partial t}=D \nabla ^2 h + K (r_+ a_+)^{m_+} |\nabla h|^{n_+} - K (r_- a_-)^{m_-} |\nabla h|^{n_-},
\end{equation}
where $D$ is the diffusion coefficient, $K>0$, $m_{\pm}>0$ and $n_{\pm}>0$ are model parameters and $r_+$/$r_-$ indicate the production rates for the respective material. The coupled nonlinear Equations (\ref{a_pm_eq_dd}) and (\ref{h_eq}) form a closed system for the interaction of counter-flowing drainage mechanisms by modifying the scalar field ($h$) with appropriate initial and boundary conditions for $h$, $a_+$ and $a_-$. 

In this work, we consider 2-dimensional and 3-dimensional domains in the shape of a rectangle or parallelepiped, respectively, with the top edge/face ($\Omega_t$) at a fixed higher value ($h=H$) compared to the bottom edge/face ($\Omega_b$) at a fixed lower value ($h=0$). The remaining side edges/faces ($\Omega_s$) follow zero Neumann boundary conditions in $h$, which provide closed boundary conditions in $a_{\pm}$. The proposed arrangement induces a directionality to the movement of the two materials in the domain with top ($\Omega_t$)  and bottom ($\Omega_b$) edges/faces functioning as the exit boundaries for $a_+$ and $a_-$, respectively. Assuming that the densities of the two materials are negligible at their upstream domain boundaries, the boundary conditions become simple and time-independent as $a_+(\Omega_b)=a_-(\Omega_t)=0$. Under such boundary conditions and the assumption of spatially uniform production rates ($r_+$ and $r_-$), the governing equations compute the counter flow of the materials across the domain (including at the boundaries where the densities are not specified i.e.,  $a_+(\Omega_t)$ and $a_-(\Omega_b)$).

From the viewpoint of the co-flowing supply and drainage mechanism (Problem II), $a_+$ represents the density of the input material in the domain, which is utilized and drained out of the domain as the output material with density $a_-$. The continuity equation for $a_-$, therefore, remains the same, with the modification in the continuity equation for the input material supplied at the boundaries, that moves with velocity $\mathbf{v_+}$ and gets consumed at the unitary rate, as 
\begin{equation}
\label{a_p_eq_sd}
\nabla \cdot \left(a_\pm\mathbf{v_\pm}\right) = \mp1,
\end{equation}
where both velocity fields, $\mathbf{v_+}$ and $\mathbf{v_-}$, follow the negative gradient of $h$ with unit speed as
\begin{equation}
\mathbf{v_\pm} = -\frac{\nabla h}{|\nabla h|} \label{v_pm_eq_sd}.
\end{equation}

Equations (\ref{h_eq}) and (\ref{a_p_eq_sd}) form a general minimalist model for the interaction of two underlying mechanisms of supply and drainage in a spatially continuous domain by modifying the scalar field ($h$), as apparent from Equation (\ref{h_eq}), where $r_+$ now represents the consumption rate of the supplied material ($a_+$). For parsimony, we here assume that the supply is consumed uniformly in space at constant rate, which is immediately disposed giving rise to a uniform and constant source of material that gets drained. More complicated patterns of supply and drainage are certainly of interest and will be investigated in the future work. The model can be analyzed from either of two discussed formulations; however, we consider the viewpoint of supply and drainage mechanisms (Problem II) from this point for the interpretation of the solutions.

The sink and source terms mathematically formalize the conceptual framework shown in Figure \ref{fig:one}, where the movement of materials carves out the preferential paths. Thus, for the 2-dimensional case the scalar field ($h$) may be viewed as an elevation field of an hypothetical landscape over which input and output materials move following Equation (\ref{v_pm_eq_sd}). As indicated by Equation (\ref{h_eq}), the accumulation of drainage material decreases the elevation that results in the formation of valleys (sink term). Conversely, the aggregation of the supply material increases the surface elevation that leads to the formation of ridges (source term). Consequently, the input material is accumulated on ridges, while the output material is concentrated in valleys. 

In the 3-dimensional case, the scalar field can be interpreted as the strength of a chemical signal that drives the movement of the materials (chemotaxis). As Equation (\ref{v_pm_eq_sd}) indicates, the concentration of the chemical signal ($h$) stimulates the migration of the materials opposite to its gradient. Vascularization of the supply and drainage networks takes place in the domain with high material density of the supply material in high-valued scalar field region and high material density of the drainage material in low-valued scalar field region due to the feedback of sink and source terms in Equation (\ref{h_eq}). The mathematical structure of the proposed model bears some resemblance with more complex models of vasculogenesis and chemotaxis \cite{chaplain2000mathematical,manoussaki2003mechanochemical}. Specifically, the core component of the model resembles minimalist versions of the well-known Keller–Segel model for chemotaxis under the negligible diffusion of biological cells \cite{keller1970initiation, hillen2007global, alber2006multiscale, hillen2009user}.

The boundary conditions play an important role in the proposed model. For Problem II, the boundary conditions for $h$ are the same as discussed for Problem I, with top ($\Omega_t$) and bottom ($\Omega_b$) edges/faces functioning now as the entry and exit boundaries for the domain, respectively. Under the assumption that no drainage material exits from $\Omega_t$ and no supply material is conveyed out of $\Omega_b$, the boundary conditions of the material densities remain simple and time-independent as $a_+(\Omega_b)=a_-(\Omega_t)=0$. Under such boundary conditions, the governing equations determine the flow of supplied and drained materials across the domain (including at the boundaries where the densities are not specified i.e.,  $a_+(\Omega_t)$ and $a_-(\Omega_b)$) under the assumption of spatially uniform consumption ($r_+$) and production rates ($r_-$).

\begin{figure}[!hbt]
\centering
\includegraphics[width=\linewidth]{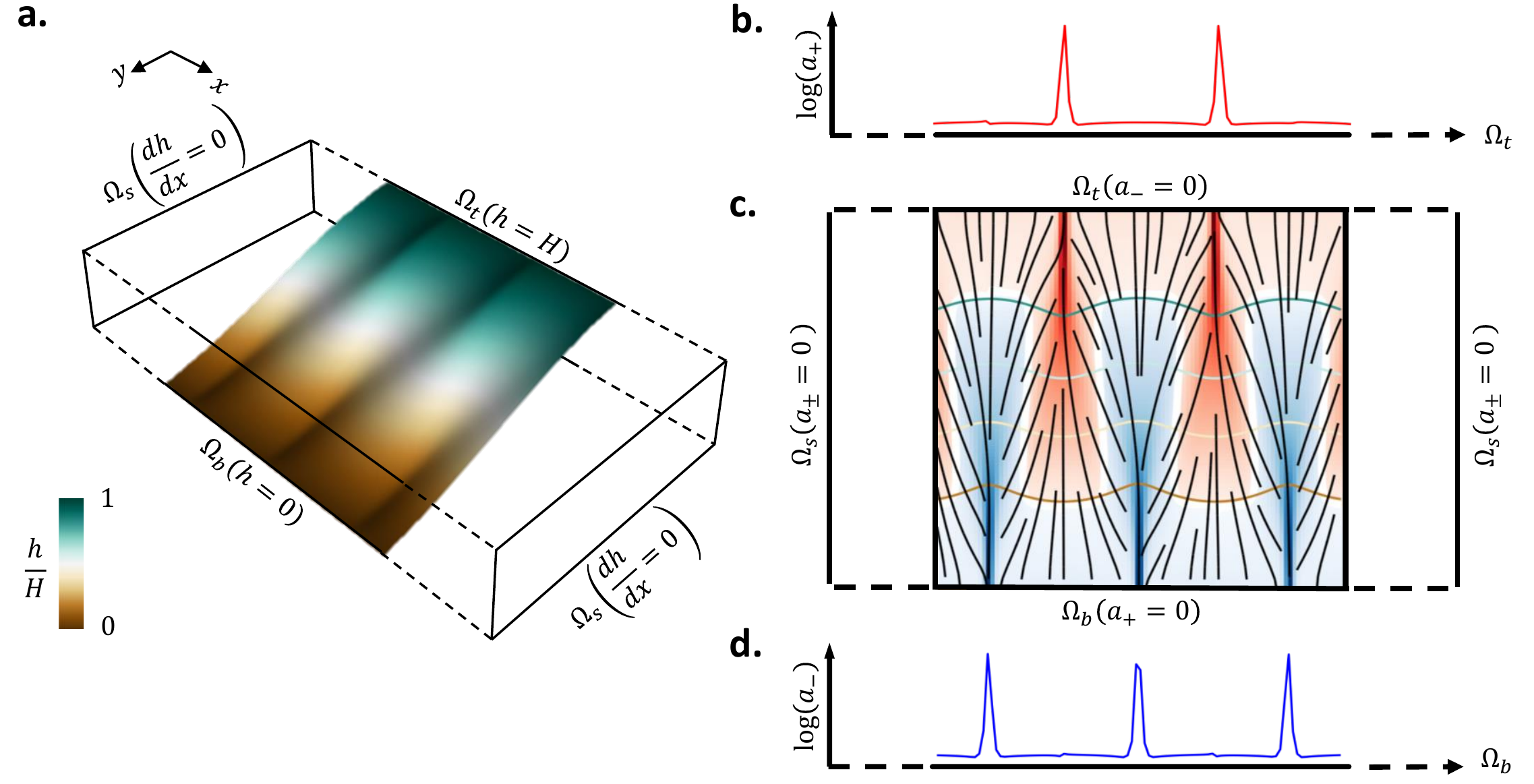}
\caption{Schematic representation of the boundary conditions used in the model. (a): Surface profile of the scalar field ($h$) in a portion of the rectangular domain near the first channel instability, where $H$ is the maximum elevation value in the domain (see Section \ref{S41} for details). Three (shallow) ridges and two (shallow) valleys can be observed in the plotted profile. (c): Boundary conditions of $a_+$ and $a_-$ counter-flow drainage problem (Problem I) and co-flow supply and drainage problem (Problem II). Two (red-colored) channels of $a_+$ and three (blue-colored) channels of $a_-$ corresponding to the ridges and valleys in panel (a) are observed. The white curve, representing the interface $a_+ = a_-$, separates the regions dominated by the either material. Four contour lines of the scalar field ($h$) are plotted along with black-colored streamlines which indicate the flow direction of the materials. (b,d): Obtained signals of $a_+$ and $a_-$ at $\Omega_t$ and $\Omega_b$, respectively, with peaks indicating channel formation at the corresponding domain boundaries.}
\label{fig:townew}
\end{figure}

It is interesting to observe that the model can be reduced to a 2-field system for the case $m_{\pm} = n_{\pm}=1$. Multiplying the continuity equations for $a_+$ and $a_-$ (Equation (\ref{a_p_eq_sd})) with $r_+$ and $r_-$, and subtracting, one can write the single equation for a new spatial field
\begin{equation}
\label{a_bar_eq_2}
a_* = \frac{r_+a_+ - r_-a_-}{r_+ + r_-}, 
\end{equation}
as 
\begin{equation}
\label{a_star_eq}
- \nabla \cdot \left(a_* \frac{\nabla h}{|\nabla h|}\right) =  -1.
\end{equation}
Equation (\ref{h_eq}) then can be re-written using Equations (\ref{a_bar_eq_2}) and (\ref{a_star_eq}) as
\begin{equation}
\label{h_eq_unit}
\frac{\partial h}{\partial t}= D \nabla ^2 h + K_* a_* |\nabla h|,
\end{equation}
where $K_* = K(r_+ + r_-)$. Equations (\ref{a_star_eq}) and (\ref{h_eq_unit}) form a 2-field equivalent formulation ($a_*,h$) to the proposed 3-field model ($a_+, a_-, h$) for unit values of the exponents in Equation (\ref{h_eq}). The achieved simplification is, in practice, only apparent as the boundary conditions of the new spatial field ($a_*$) for the reduced model require the knowledge of $a_+$ and $a_-$ in advance to obtain the same solution given by the 3-field model with the time-independent boundary conditions. The reader is referred to \ref{2field}, where this problem of the boundary condition for the 2-field model is discussed in detail for simulation results in the 2-dimensional case.

\subsection{Non-dimensionalization}
\label{S23}

For a typical value $H$ of the scalar field and a typical length scale of the domain $L$, the following dimensionless quantities are established: $\hat{h} = \frac{h}{H}$, $\hat{a}_{+} = \frac{a_+}{L}$, $\hat{a}_{-} = \frac{a_-}{L}$, $\hat{t} = \frac{L^2}{D}$, $\hat{x} = \frac{x}{L}$ and $\hat{y} = \frac{y}{L}$. Using these quantities, Equations (\ref{h_eq}) and (\ref{a_p_eq_sd}) can be written in dimensionless form, 
\begin{equation}
\label{h_eq_nd}
\frac{\partial \hat{h}}{\partial \hat{t}} = \hat{\nabla} ^2 \hat{h}
+ \mathcal{C_{I_+}} \hat{a}_{+}^{m_+} |\hat{\nabla} \hat{h}|^{n_+}
- \mathcal{C_{I_-}} \hat{a}_{-}^{m_-} |\hat{\nabla} \hat{h}|^{n_-},
\end{equation}
\begin{equation}
\label{v_pm_eq_sd_nd} 
- \hat{\nabla} \cdot \left( \hat{a}_{\pm} \frac{\hat{\nabla} \hat{h}}{|\hat{\nabla} \hat{h}|}\right) =  \mp 1,
\end{equation}
where
\begin{equation}
\label{chi}
\mathcal{C_{I_+}} = \frac{K r_+^{m_+} L^{2 + m_{+}-n_{+}}}{DH^{1-n_+}}, \mathcal{C_{I_-}} = \frac{K r_-^{m_-} L^{2 + m_{-}-n_{-}}}{DH^{1-n_-}}.
\end{equation}

This shows that the overall behavior of the system can be described by the two `channelization indices', $\mathcal{C_{I_+}}$ and $\mathcal{C_{I_-}}$. For a constant value of exponents in  Equation (\ref{h_eq_nd}), an increase in the value of $\mathcal{C_{I_+}}$ by high consumption rate, $r_+$, enhances the feedback of the source term. On the other hand, a rise in the value of $\mathcal{C_{I_-}}$ by high production rate, $r_-$, strengthens the feedback of the sink term, keeping all other factors the same. This mechanism results in a correlation of the density of the two materials to the value of the scalar field at steady-state which can be visualized by looking at the level set $L_c(h)$ of the scalar field $h$ for a constant value $c$. High density of input/output material accruing on the different level sets of the scalar field is shown in the steady-state solutions of the 2-dimensional and 3-dimensional cases (Section \ref{S4}).

\section{\label{S3} Closed-form solution}

At steady-state, the closed-form solution can be obtained for the case where diffusion in Equation (\ref{h_eq_nd}) inhibits the instability formation in the scalar field. In the 2-dimensional case, it can be visualized as the smooth elevation field in a semi-infinite domain where the top edge, which is at a fixed higher elevation ($H$) compared to the bottom edge, is separated by the distance $L$ from the bottom edge. This situation is analogous to the flow of two materials before vascularization across two infinite parallel plates placed at a finite distance $L$ in 3-dimension, with a fixed high chemical signal's strength ($H$) at the top face compared to the fixed zero chemical signal’s strength at the bottom face, which drives the flow of the materials.

Assuming that the scalar field ($\hat{h}$) decreases monotonically in the 1D transect, Equation (\ref{v_pm_eq_sd_nd}) can be solved with the boundary conditions $\hat{a_+}(\hat{y}=1) = \hat{a}_-
(\hat{y}=0) = 0$ to obtain $\hat{a_+} = (1 - \hat{y})$ and $\hat{a_-} = \hat{y}$. For the case of $m_{\pm} = n_{\pm} = 1$, substituting the expressions for $\hat{a_+}$ and $\hat{a_-}$ in Equation (\ref{h_eq_nd}) at steady-state can be written as
\begin{equation}
\label{h_eq_nd_ss}
\hat{h}'' + \mathcal{C_{I_-}} \hat{y} h' - \mathcal{C_{I_+}} (1-\hat{y}) \hat{h}'= 0.
\end{equation}

Solving Equation (\ref{h_eq_nd_ss}) gives
\begin{equation}
\label{h_sol}
\hat{h} = \frac{erf\Big(\frac{\mathcal{C_{I_-}}}{\sqrt{2(\mathcal{C_{I_-}} + \mathcal{C_{I_+}})}}\Big) - 
erf\Big(\frac{\mathcal{C_{I_-}}\hat{y} - \mathcal{C_{I_+}}(1-\hat{y})}{\sqrt{2(\mathcal{C_{I_+}} +  \mathcal{C_{I_-}})}}\Big)
}{erf\Big(\frac{\mathcal{C_{I_-}}}{\sqrt{2(\mathcal{C_{I_-}} + \mathcal{C_{I_+}})}}\Big) + 
erf\Big(\frac{\mathcal{C_{I_+}}}{\sqrt{2(\mathcal{C_{I_-}} + \mathcal{C_{I_+}})}}\Big)}
\end{equation}

\begin{eqnarray}
\label{h_dash_sol}
|\hat{h'}| = \frac{e^{-\frac{(\mathcal{C_{I_-}}\hat{y} - \mathcal{C_{I_+}} (1-\hat{y}))^2}{2(\mathcal{C_{I_-}} + \mathcal{C_{I_+}})}}
}{erf\Big(\frac{\mathcal{C_{I_-}}}{\sqrt{2(\mathcal{C_{I_-}} + \mathcal{C_{I_+}})}}\Big) + 
erf\Big(\frac{\mathcal{C_{I_+}}}{\sqrt{2(\mathcal{C_{I_-}} + \mathcal{C_{I_+}})}}\Big)}\times
\sqrt{\frac{2(\mathcal{C_{I_-}}+\mathcal{C_{I_+}})}{\pi}}.
\end{eqnarray}

\begin{figure}[!hbt]
\centering
\includegraphics[width=\linewidth]{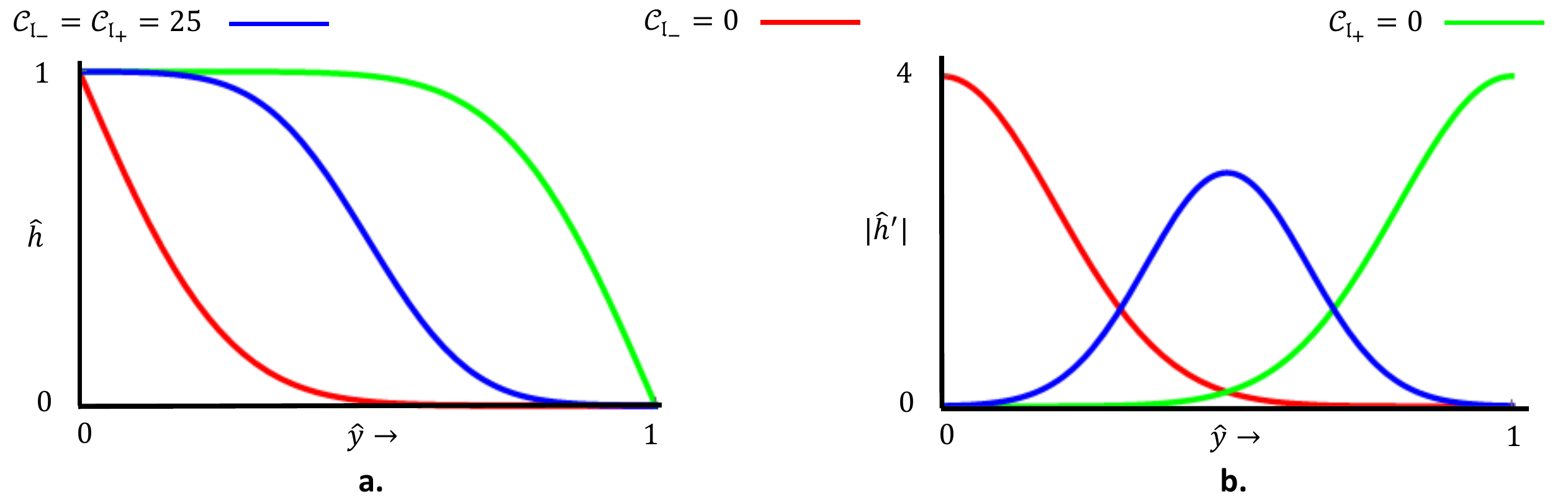}
\caption{(a,b): Steady-state solutions given by Equation (\ref{h_sol}) and (\ref{h_dash_sol}) for three cases of $\mathcal{C_{I_+}} = 0$, $\mathcal{C_{I_-}}=0$ and $\mathcal{C_{I_\pm}} = 25$.}
\label{fig:three}
\end{figure}

Smooth profiles using Equation (\ref{h_sol}) and the corresponding slope variations following Equation (\ref{h_dash_sol}) for $\mathcal{C_{I_+}} = 0$, $\mathcal{C_{I_-}}=0$ and $\mathcal{C_{I_+}}= \mathcal{C_{I_-}} = 25$ are displayed in Figure \ref{fig:three} (a,b). As expected, for $\mathcal{C_{I_-}} = 0$, the contribution from the nonlinear sink term goes away and the surface attains a higher profile compared to the case for $\mathcal{C_{I_+}} = 0$.

\section{ Numerical solutions}
\label{S4}

Numerical experiments are started for the 2-dimensional and 3-dimensional cases with a linear initial condition  containing a small amount of random spatial noise. We limit our discussion to the case with unity exponents ($m_{\pm} = n_{\pm} =1$), and utilize the efficient algorithm presented in \cite{2019arXiv190903176A} to update the scalar field $h$ over the entire domain until the steady-state is reached. The fundamental concept behind this algorithm is inspired from the notion of a flow network of the material over the entire scalar field, which is traversed in a way to make the matrix system upper/lower triangular for the efficient implicit computation. The accuracy of the numerical algorithm has been carefully tested for the case of drainage-network evolution model for the natural landscape against analytical solutions in non-channelized/vascularized conditions, as well as against analytical results of the onset of linear stability analysis \cite{bonetti2020channelization, 2019arXiv190903176A} and with exact mean field solutions obtained in the condition of fully channelized/vascularized regime \cite{2019arXiv190903176A}. We refer to these references for further details.

\subsection{\label{S41} 2-dimensional case}
\subsubsection{\label{S411} Code Verification}
\begin{figure}[!hbt]
\centering
\includegraphics[width=0.88\linewidth]{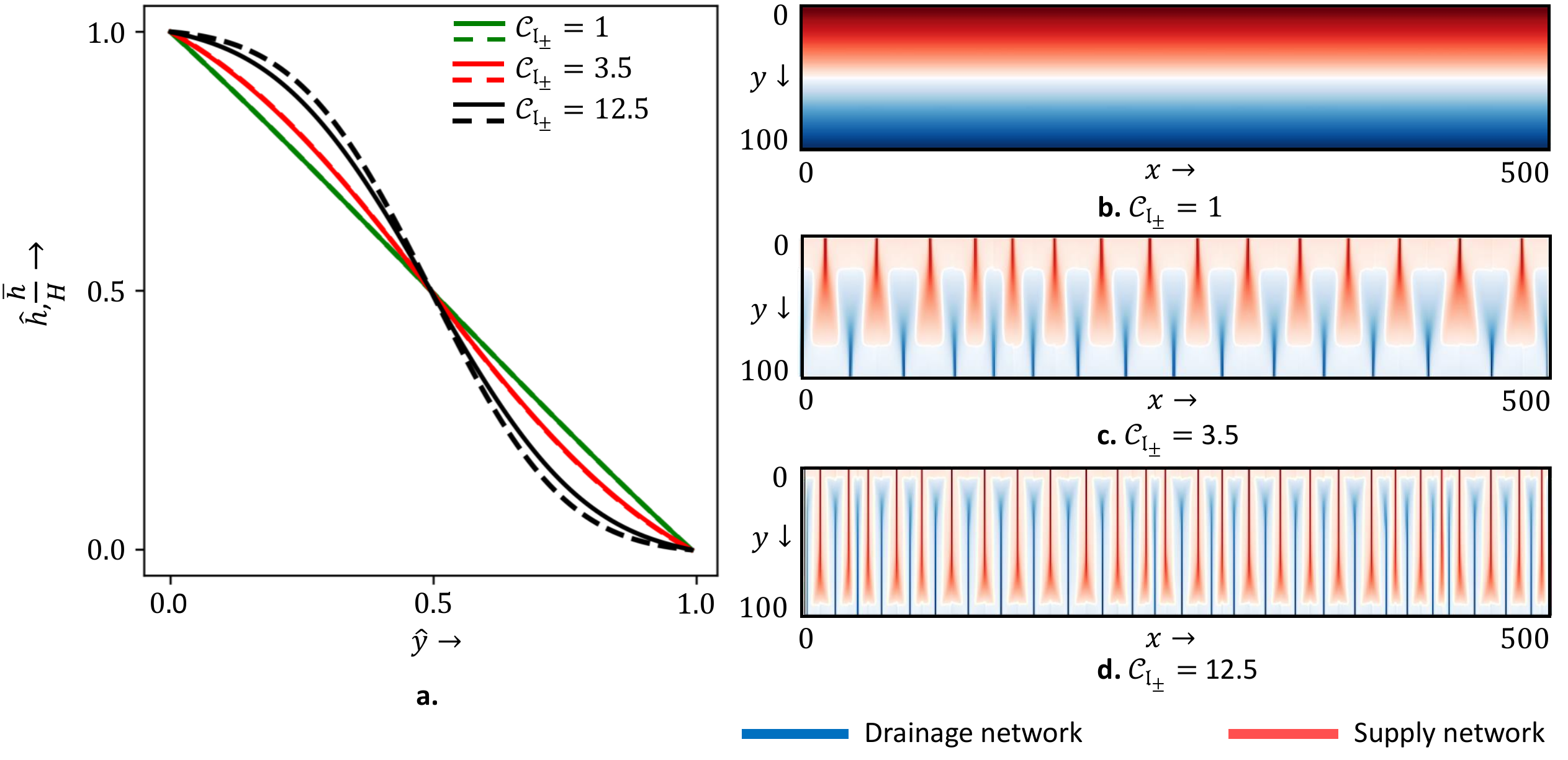}
\caption{ (a): Solid lines represent the computed mean surface profile ($\bar{h}$) along the length for a rectangular domain (width = 100, length = 500) for $\mathcal{C_{I_{\pm}}} = 1$, $3.5$ and $12.5$ compared to the steady-state closed-form solutions given by Equation (\ref{h_sol}) as dashed lines. (b,c,d): Simulation results for the accumulation of $a_+$ (red-colored) and $a_-$ (blue-colored) for $\mathcal{C_{I_{\pm}}} = 1$, $3.5$ and $12.5$. The white curve represents the interface $a_+ = a_-$, separating the two regions dominated by the either material.}
\label{fig:four}
\end{figure}

We first simulate a rectangular domain with high aspect ratio (length  = 500, width = 100) and compare the mean elevation profile along the length for varying values of $\mathcal{C_{I_{\pm}}}$ to verify the implemented code. The first channelization in the domain occurs at $\mathcal{C_{I_{\pm}}} = 3.5$. The closed-form solution is applicable for the cases when the field $h$ is smooth enough (no channelization). The mean surface profile starts deviating from the closed-form solution for $\mathcal{C_{I_{\pm}}} \geq 3.5$ (Figure \ref{fig:four}(a)) due to channel formation. This is apparent from the accumulation plot of $a_+$ and $a_-$, where the white curve represents the interface for $a_- = a_+$. For $\mathcal{C_{I_{\pm}}} = 1$, the interface is a straight line, while for $\mathcal{C_{I_{\pm}}} = 3.5$ (the onset of first channelization) and $12.5$, the interface becomes a curve due to the emergence of channels in supply and drainage networks (Figure \ref{fig:four} (b,c,d)). 

\subsubsection{\label{S412} Effect of individual factors}

In this numerical experiment, we focus on the impact of individual factors on the coupled-network formations. Varying the relative rate of consumption ($r_+$) of the supplied material versus the generation rate ($r_-$) of the drained material affects the feedback on the scalar field, which in turn affects the structure of the coupled networks. Having all other parts parameters the same, the variation in these rates can be expressed as changing the values of non-dimensional indices $\mathcal{C_{I_{+}}}$ and $\mathcal{C_{I_{-}}}$.

\begin{figure}[!hbt]
\includegraphics[width=\linewidth]{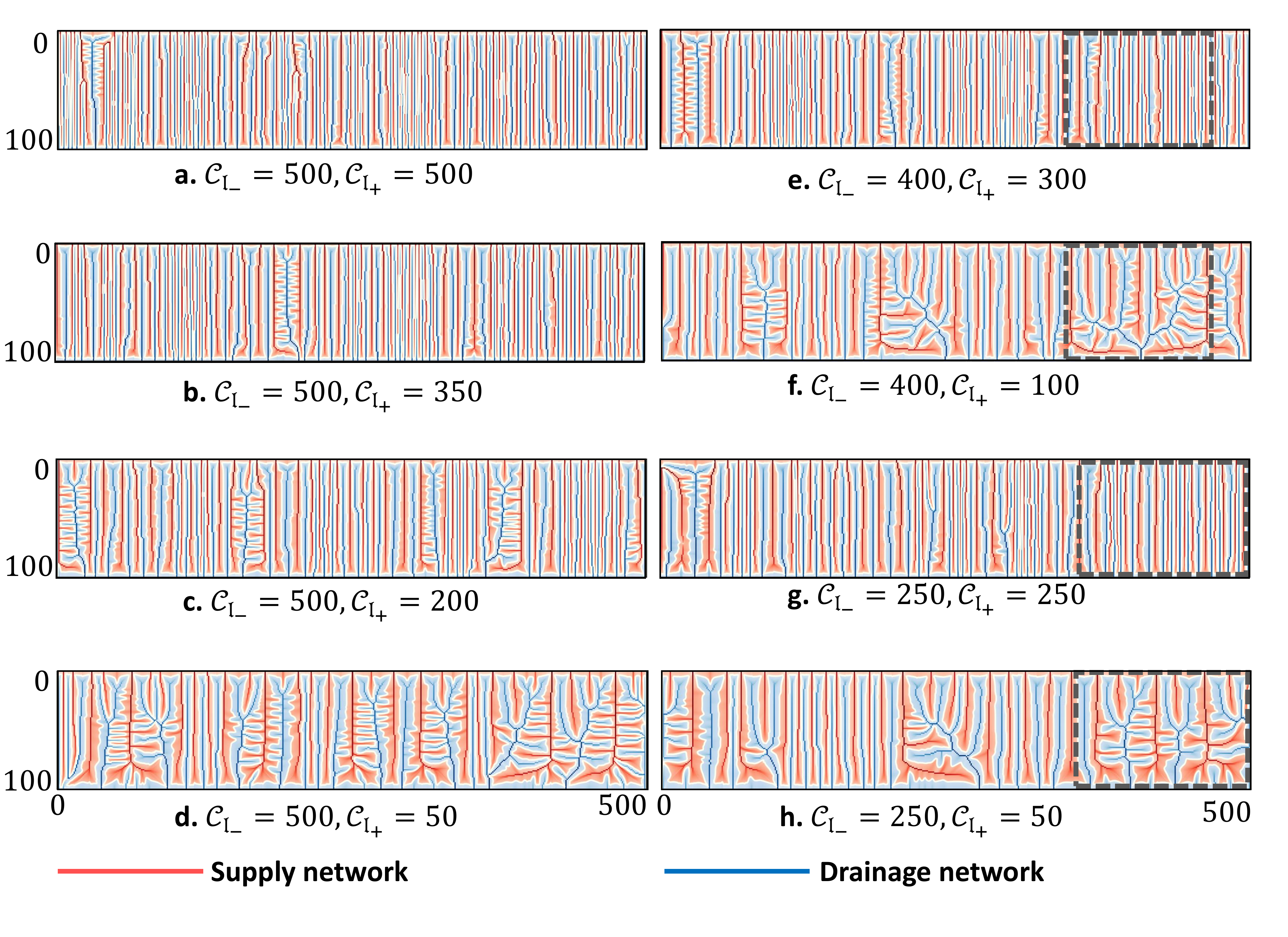}
\caption{\label{fig:five} Simulation results for various values of $\mathcal{C_{I_+}}$ and $\mathcal{C_{I_-}}$ in a rectangular domain (width = 100, length = 500). The accumulation of the input material $a_+$ is represented in red (highlighting $a_+>a_-$), while the accumulation of the output material $a_-$ is shown in blue (highlighting $a_+<a_-$). The white curve represents the interface $a_+ = a_-$ and separates region dominated by the input material from the output material. The grey-colored boxes on the panels e, f, g and h indicate the regions of the domain for which 3-dimensional surface profiles of the elevation field ($h$) are shown in Figure \ref{fig:six}.}
\end{figure}

We study the extent and spatial patterns of $a_-$ and $a_+$ for 55 cases with $\mathcal{C_{I_{\pm}}} \in [50, 500]$ and $\mathcal{C_{I_+}} \leq \mathcal{C_{I_-}}$. Figure \ref{fig:five} presents simulation results, where the supply network is represented in red (highlighting high density region of $a_+$, i.e., $a_+>a_-$) and the drainage network is shown in blue color (accentuating high density region of $a_-$, i.e., $a_->a_+$) with the white curve representing the interface $a_+=a_-$. These spatial networks that evolve for various values of $\mathcal{C_{I_{\pm}}}$ are quite distinctive, indicating the role of absolute as well as relative values of $\mathcal{C_{I_{+}}}$ and $\mathcal{C_{I_{-}}}$ on the overall pattern formation. Panels (a,b) of Figure \ref{fig:five} display the plots where the number of channels of the supply and drainage network is high, with mostly straight channels and very little branching. Panels (c,e,g) present the plots where comparatively less number of channels are observed with more branching. Panels (d,f,h) show the plots where maximum branching is observed with curved channels compared to previous other cases.

\begin{figure}[!hbt]
\includegraphics[width=\linewidth]{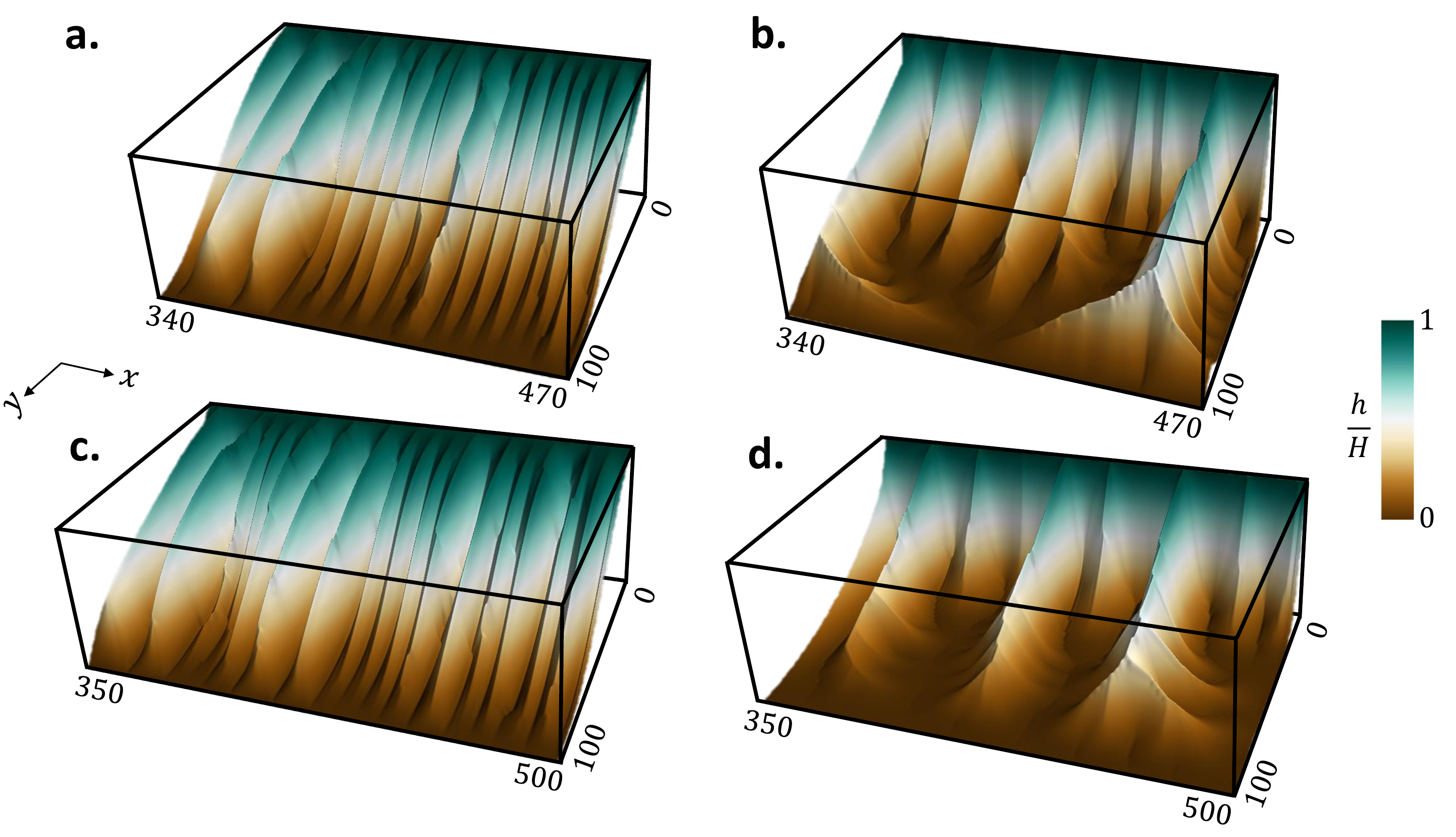}
\caption{\label{fig:six} 3-dimensional surface profiles of $h$ for the regions highlighted in Figure (\ref{fig:five}). (a): $\mathcal{C_{I_-}} = 400$, $\mathcal{C_{I_+}} = 300$. (b): $\mathcal{C_{I_-}} = 400$, $\mathcal{C_{I_+}} = 100$. (c): $\mathcal{C_{I_-}} = 250$, $\mathcal{C_{I_+}} = 250$. (d): $\mathcal{C_{I_-}} = 250$, $\mathcal{C_{I_+}} = 50$. Comparable values of $\mathcal{C_{I_+}}$ and $\mathcal{C_{I_-}}$ result in the formation of shallow ridges and valleys with less branching, while the disproportionate values of $\mathcal{C_{I_+}}$ and $\mathcal{C_{I_-}}$ result in wide and more branched valleys with sharp ridges.}
\end{figure}

The relative strength of $\mathcal{C_{I_+}}$ and $\mathcal{C_{I_-}}$ affecting the shape of the surface ($h$) and hence, the spatial patterns of both networks is apparent from Figure \ref{fig:six}. The figure displays the 3-dimensional surface plots of $h$ from the selected regions in Figure \ref{fig:five}. Panels (a) and (c) display the surface plots where the comparable opposing strength of $a_+$ and $a_-$ results in the formation of shallow ridge and valley patterns. Panels (b) and (d) show the surface plot for the branched region of $\mathcal{C_{I_+}} = 100$, $\mathcal{C_{I_-}} = 400$ and $\mathcal{C_{I_+}} = 50$, $\mathcal{C_{I_-}} = 250$ where high value of $\mathcal{C_{I_-}}$ compared to $\mathcal{C_{I_+}}$ results in branched channels of the networks with wide valleys and thin ridges at the steady-state.

The interplay between model parameters and boundary conditions becomes apparent when considering panels f and g. These cases have the same total $\mathcal{C_{I_+}}+ \mathcal{C_{I_-}}$ for the 2-field model, and thus, both solutions satisfy the same differential equations (\ref{a_star_eq}) and (\ref{h_eq_unit}). Nevertheless, as is apparent in Figure \ref{fig:five}, resulting branched structures are vastly different. This shows that the non-trivial boundary conditions allowed by the three-field model (as opposed two-field model) influences the solution throughout the domain, both quantitatively and qualitatively. A full discussion of this is included in \ref{2field}.

\begin{figure}[!hbt]
\centering
\includegraphics[width=\linewidth]{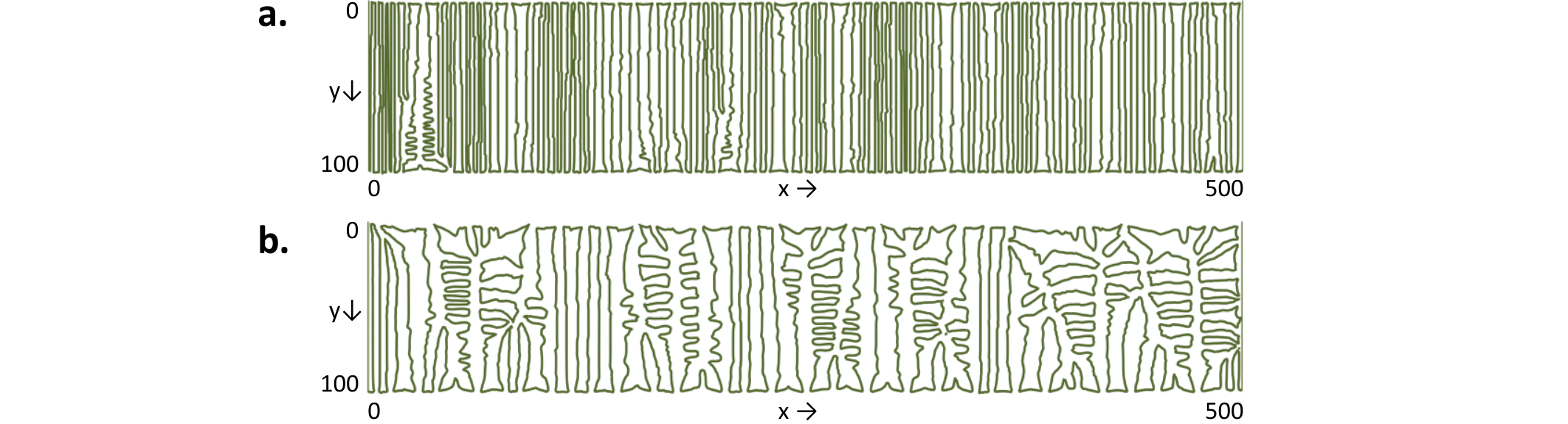}
\caption{Plot of the interface $a_+ = a_-$ for ($\mathcal{C_{I_+}} = 500$, $\mathcal{C_{I_-}} = 500$) and ($\mathcal{C_{I_+}} = 50$, $\mathcal{C_{I_-}} = 500$).}
\label{fig:seven}
\end{figure}

The variety of patterns in above cases can be explained by the structure of Equation (\ref{h_eq_nd}) for the 3-field model. Increasing values of $\mathcal{C_{I_{\pm}}}$ hike the tendency of $a_+$/$a_-$ to mold the surface as ridge and valley respectively. For very high and comparable values of $\mathcal{C_{I_-}}$ and $\mathcal{C_{I_+}}$, the primary channels get stuck, can not coalesce to form branched channels. Therefore, the number of main channels (channels originating from the boundaries) increases, which increases the length of the interface $a_+ = a_-$. Reducing the value of $\mathcal{C_{I_+}}$ with respect to $\mathcal{C_{I_-}}$ results in more branching as channels of $a_-$ dominate the space and coalesce together to form branched patterns. For $\mathcal{C_{I_-}} = 500$, changing $\mathcal{C_{I_+}} = 50$ to $\mathcal{C_{I_+}} = 500$ increases the length of interface ($a_+ = a_-$) as shown in Figure \ref{fig:seven}.

\begin{figure}[!hbt]
\centering
\includegraphics[width=\linewidth]{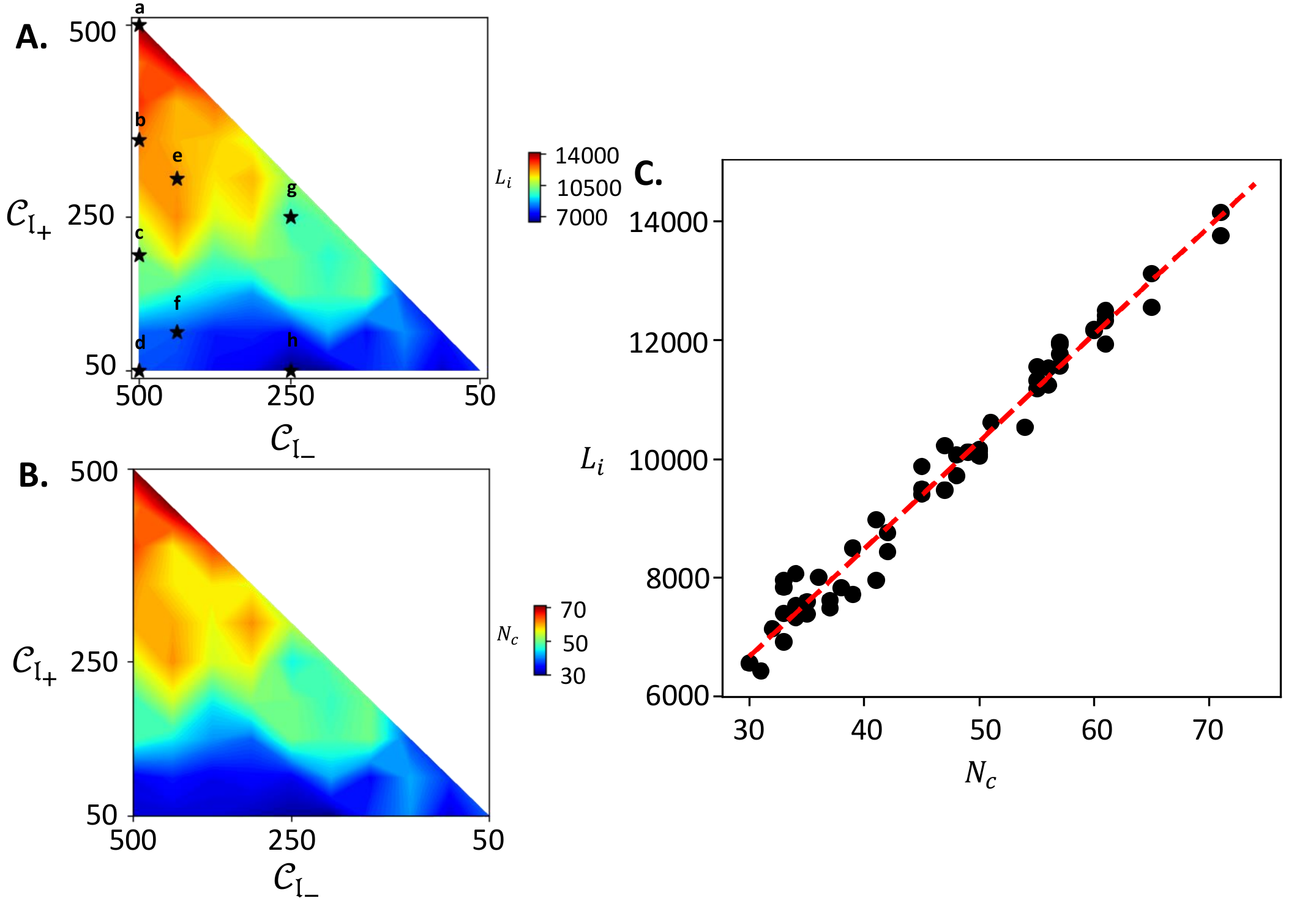}
\caption{(A): Color-plot of the interface length, $L_i$, for various values of $\mathcal{C_{I_\pm}} \in (50, 100, 150,... 500)$. (a-h) points indicate the cases shown in Figure \ref{fig:five}. (B): Color-plot of the number of main channels, $N_c$, for various values of $\mathcal{C_{I_+}}$ and $\mathcal{C_{I_-}} \in (50,100, 150,... 500)$. (C): Scatter plot of $L_c$ vs $N_c$ with the best-fit line (Correlation coefficient $r = 0.988$).}
\label{fig:eight}
\end{figure}

We plot the length of the interface $a_+=a_-$ (denoted by $L_i$) for various values of $\mathcal{C_{I_{\pm}}}$. High values of $L_i$ occur for large and comparable values of $\mathcal{C_{I_{\pm}}}$ which is shown as the red-colored region in Figure \ref{fig:eight}(A). Conversely, for disproportionate values of $\mathcal{C_{I_{\pm}}}$, the interface length ($L_i$) is relatively smaller as shown in the blue-colored region in Figure \ref{fig:eight}(A). We define a quantity $N_{c}$ which refers to the maximum number of either main supply or drainage channels of length greater than the half of the width of the domain (50 in this case) originating from the boundaries of the domain. $N_{c}$ is plotted for 55 cases of various values of $\mathcal{C_{I_{\pm}}}$ as Figure \ref{fig:eight}(B), which looks similar to the plot of $L_i$ as expected. High values of $N_c$ occur for large and similar values of $\mathcal{C_{I_{\pm}}}$ again shown as the red-colored region. More branching results in less number of main channels for $\mathcal{C_{I_+}} << \mathcal{C_{I_-}}$, as indicated by the blue-colored in Figure \ref{fig:eight}(B). The scatter plot of $L_i$ versus $N_c$ with best-fit line having correlation coefficient $r = 0.988$ reconfirms the close relationship between number of main channels and the interface length (Figure \ref{fig:eight}(C)). 

The simulation results shown in Figure \ref{fig:five} can be mapped to different regions in the color-plot of the contour length $L_c$. Panels (a,b) in Figure \ref{fig:five} belong to the red-colored region in Figure \ref{fig:eight}(A), where a high density of nearly unswerving main channels with a few offshoots is observed. For this reason, we refer to it as the congested region. Panels (d,f,h) in Figure \ref{fig:five} exhibit the plots for the blue (branched) area in Figure \ref{fig:eight}(A), with heavily branched channels. Panels (c,e,g) in Figure \ref{fig:five} display the plots from the yellow/green (transient) region, which lies within these two extremes.

\subsection{3-dimensional case}
\label{42}

\begin{figure}[!htb]
\includegraphics[width=\linewidth]{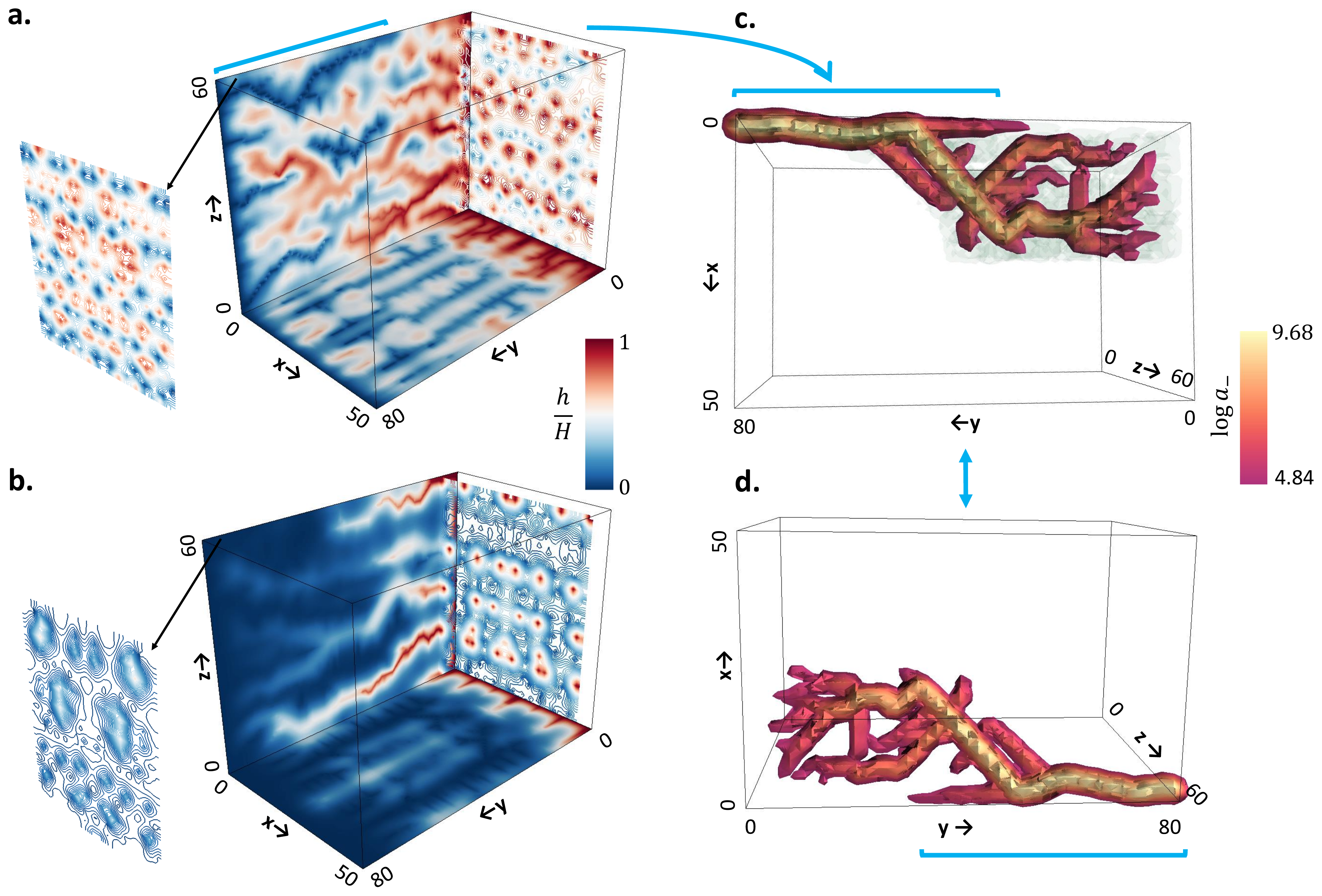}
\caption{\label{fig:nine} Simulation results for the 3-dimensional domain  ($x = 50,y = 80,z = 60$). Contour plot presenting the scalar $h$ at the side boundaries and at the two cross-sections along the $y$-axis near the respective ends (top and bottom face) for (a): $\mathcal{C_{I_+}} = 1000$ and $\mathcal{C_{I_-}} = 1000$. (b): $\mathcal{C_{I_+}} = 200$ and $\mathcal{C_{I_-}} = 1000$. (c,d): The largest drainage conduit extracted in the domain for the case of $\mathcal{C_{I_{\pm}}} = 1000$ with light green colored haze (in panel c) indicating the points in the domain from which the flow of drainage material is received in the conduit.}
\end{figure}

We apply the proposed model to a 3-dimensional domain for a parallelepiped ($x = 50,y = 80,z = 60$), where $h$ now refers to a density field (can be viewed as a chemical signal's strength). There are fixed boundary conditions for two faces ($h(x,0,z)= H = 10$ and $h(x,80,z)= 0$) and zero Neumann boundary conditions at the remaining faces. $a_-$ is zero at $h(x,0,z)= H = 10$ and $a_+$ is zero at $h(x,80,z)= 0$, with closed boundary conditions in $a_{\pm}$ for the remaining four faces. We explore two cases keeping $\mathcal{C_{I_{-}}} = 1000$, while changing $\mathcal{C_{I_{-}}}$ from $1000$ to $200$. The simulation results are shown in Figure \ref{fig:nine}, where the contour plots for the field $h$ are drawn on the two side faces (closed boundary conditions in $a_\pm$) along with contour line plots for the two cross-sections near the faces along $y$-axis.

The steady-state solutions for the 3-dimensional cases agree with the patterns witnessed in the 2-dimensional results. As shown in Figure \ref{fig:nine}(a), a large number of red-colored contour curves (high density region of $h$) in cross-section near the face $h(x,0,z)$ remain in cross-section near the face $h(x,80,z)$, which is dominated by the blue-colored contour curves (low density region of $h$). This pattern for $\mathcal{C_{I_\pm}}=1000$ resembles the equal strength of shallow ridges and valleys in the 2-dimensional case. We display the largest drainage conduit from the steady-state solution in Figure \ref{fig:nine}(c,d), where green-colored haze in panel c indicates the points in the domain from which the flow is collected in the given conduit.

Similarly, the contour patterns for $\mathcal{C_{I_+}} = 200$ and $\mathcal{C_{I_-}} = 1000$ on the faces resemble the thin ridges and wide valleys obtained in the 2-dimensional branched case when the values of $\mathcal{C_{I_{+}}}$ and $\mathcal{C_{I_{-}}}$ are disproportionate (Figure \ref{fig:nine}(b)). This is apparent as the tiny red-colored contour curves (high density region of $h$) in the cross-section near the face $h(x,0,z)$ vanish in the cross-section near the face $h(x,80,z)$ dominated by the blue-colored contour curves (low density region of $h$). This parallels the thin ridges that start from the fixed elevation end ($h=H$) and disappear near the fixed elevation end ($h=0$) surrounded by deep and wide valleys, leading to the branched supply and drainage networks.

\section{\label{S5} Conclusions}

The minimalist model developed in this work leads to the formation of spatial patterns of combined supply and drainage networks in a continuous domain, whereby the corrugations of a mediating scalar field, $h$, cleave these competing networks in the same continuous domain. A channelization index ($\mathcal{C_{I_\pm}}$) corresponding to each material governs the relative intensity of the branching of these networks and the instability in the profile of $h$. The crucial role of the boundary condition for these coupled PDEs is particularly evident when reducing the presented 3-field model to a 2-field model for unit values of the exponents in source and sink terms, as the achieved simplification in the number of equations entails a complication in the boundary conditions, which is necessary to solve the same co-existing supply and drainage networks of the 3-field model. 

While we limit our discussion here to unit exponents of the source and sink terms in Equation (\ref{h_eq_nd}), the solutions for non-unitary values of the exponents have qualitatively similar features and reflect an analogous spectrum of branched versus congested regime after the first channelization for a different range of $\mathcal{C_{I_\pm}}$. However, the specific results do depend on these nonlinearities and the source and sink terms: future work will be devoted to adjusting them to cater to specific applications, as has been done in various other models, such as the minimalist versions of the well-known Keller–Segel model for chemotaxis \cite{hillen2009user,keller1970initiation}, mechanochemical models of angiogenesis and vasculogenesis \cite{manoussaki1996mechanical,manoussaki2003mechanochemical}. 

From the numerical point of view, the employed algorithm decreases the time complexity of the implicit solver by making the matrix system upper/lower triangular. This has been a vital improvement in 2-dimensional cases, where the space complexity of the algorithm is not an issue \cite{2019arXiv190903176A}. However, the simulations in the 3-dimensional domain require a large amount of memory space compared to the 2-dimensional cases due to the increased input size of nodes and the corresponding auxiliary space utilized by the algorithm during the execution. This increases the overall computational cost of the simulations in the 3-dimensional cases. A part of the future work, therefore, is to reduce the space complexity of the numerical solver so that the coupled patterns for a 3-dimensional domain can be analyzed in more depth.

\section*{Acknowledgment}
The authors acknowledge support from the US National Science Foundation (NSF) grants EAR-1331846 and EAR-1338694, and BP through the Carbon Mitigation Initiative (CMI) at Princeton University. A.P. and M.H. also acknowledge the support from the Princeton Institute for International and Regional Studies (PIIRS) and the Princeton Environmental Institute (PEI). J. M. N. was supported in part through Norwegian Research Council grant number 250223.

The authors are pleased to acknowledge that the simulations presented in this article were performed on computational resources managed and supported by Princeton Research Computing, a consortium of groups including the Princeton Institute for Computational Science and Engineering (PICSciE) and the Office of Information Technology's High Performance Computing Center and Visualization Laboratory at Princeton University.

The code used for the simulations is available at \url{https://github.com/ShashankAnand1996/Supply-Drainage}.

\bibliographystyle{elsarticle-num}
\bibliography{reference}

\appendix
\section{2-field model}
\label{2field}

In Section \ref{S22}, we show that the original 3-field model can be reduced to a 2-field model consisting of Equations (\ref{a_star_eq}) and (\ref{h_eq_unit}) for the new spatial field, $a_*$, and the scalar field, $h$, under the assumption of unit exponents of the source and sink term in Equation (\ref{h_eq}). These equations form a closed system where the dynamics of $h$ depends on the parameter $K_*$, which is determined by the summation of $r_+$ and $r_-$ only, instead of the two channelization indices that are defined for the 3-field model.

\begin{figure}[!hbt]
\includegraphics[width=\linewidth]{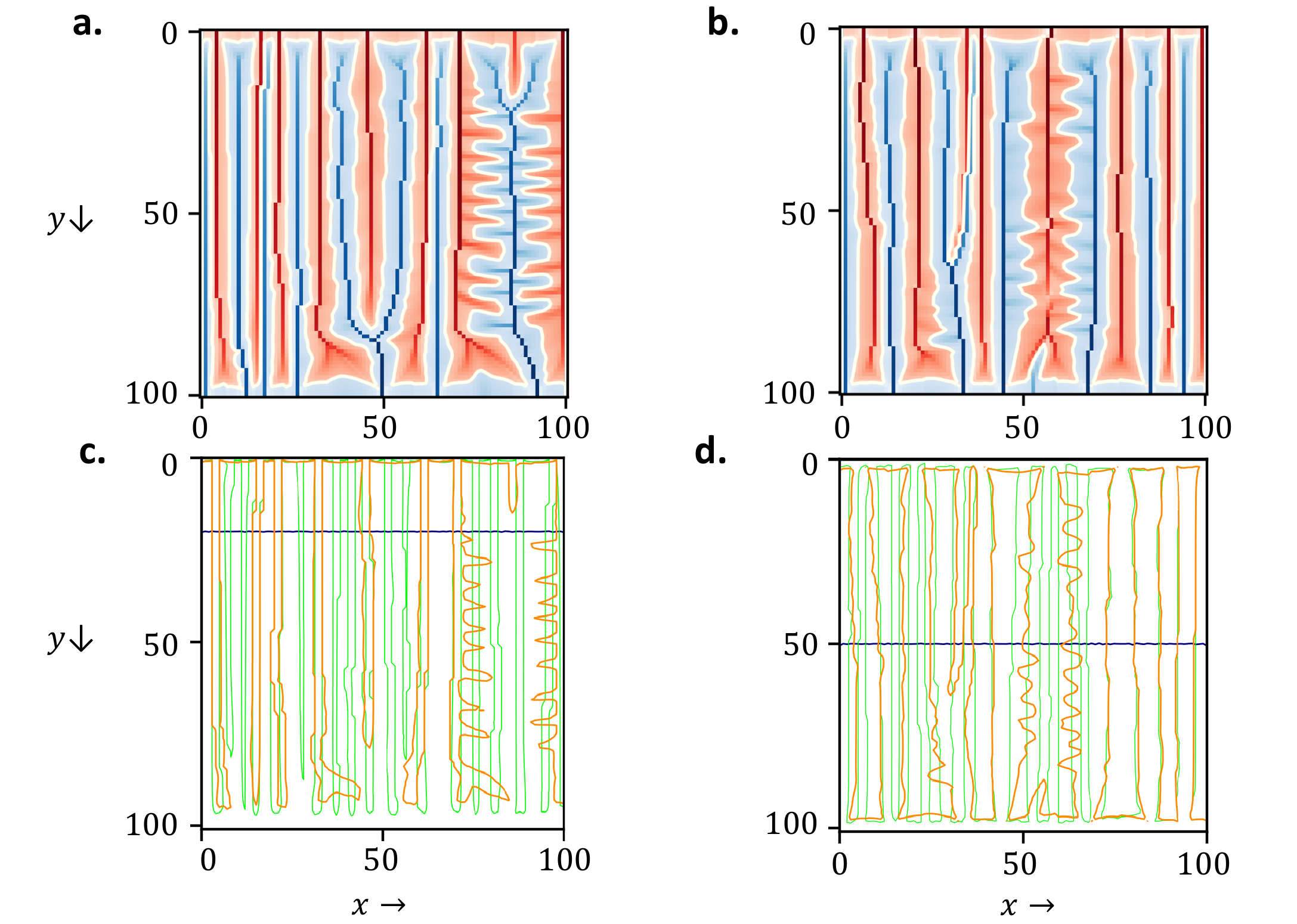}
\caption{(a,b): Steady-state solutions for $\mathcal{C_{I_+}} = 100$, $\mathcal{C_{I_-}} = 400$, and $\mathcal{C_{I_+}} = 250$, $\mathcal{C_{I_-}} = 250$, respectively. The accumulation of $a_+$ is in red ($a_+>a_-$), the accumulation of $a_-$ is in blue ($a_+<a_-$) and the white curve is interface $a_+ = a_-$. (c,d): $a_* = 0$ at $t = 0$ (blue), $t= \text{intermediate}$ (green) and $t=\text{steady state}$ (orange) for cases (a) and (b), respectively.}
\label{fig:ten}
\end{figure}

We discuss here the dependency of complex boundary conditions of $a_*$ on the solution of spatial fields $a_+$ and $a_-$ by presenting steady-state solutions for a 2-dimensional square domain with top edge ($y=0$) at fixed high elevation ($H= 10$) and bottom edge ($y=L=100$) at fixed zero elevation, with zero Neumann boundary conditions on the side edges. With the same values of $D = 10^{-3}$, $K = 10^{-5}$ and $(r_+ + r_-) = 5$, two cases were simulated as $r_+ = 1$, $r_- = 4$ ($\mathcal{C_{I_+}} = 100$, $\mathcal{C_{I_-}} = 400$), and $r_{\pm} = 2.5$ ($\mathcal{C_{I_{\pm}}} = 250$). Figures \ref{fig:ten}(a,b) show the plots of steady-state supply and drainage material densities ($a_+$ and $a_-$) for the two cases. The difference in the obtained supply and drainage networks can be interpreted as the role of different values of $\mathcal{C_{I_{\pm}}}$ in the 3-field model. However, the two cases correspond to the same value of $K_* = 5 \times 10^{-5}$ for the 2-field model, which indicates the crucial role of time-dependent boundary condition of $a_*$ on the obtained supply and drainage networks.

For the 3-field model, time-independent boundary condition for $a_+$ is well defined, with $a_+ = 0$ at the bottom edge ($h= 0$) for the both cases. Similarly, the boundary condition for $a_-$ is fixed in time throughout the simulations with $a_- = 0$ at the top edge of the square domain. For the 2-field model, the boundary condition for $a_*$ in the two cases is different and is defined by specifying individual values of $r_+$ and $r_-$ initially, as shown in Figure \ref{fig:ten}(c,d). The blue curves for both cases, representing $a_* = 0$, vary in time, indicating the contribution of time-dependent boundary condition of $a_*$ on the simulation results.

\begin{figure}[!hbt]
\includegraphics[width=\linewidth]{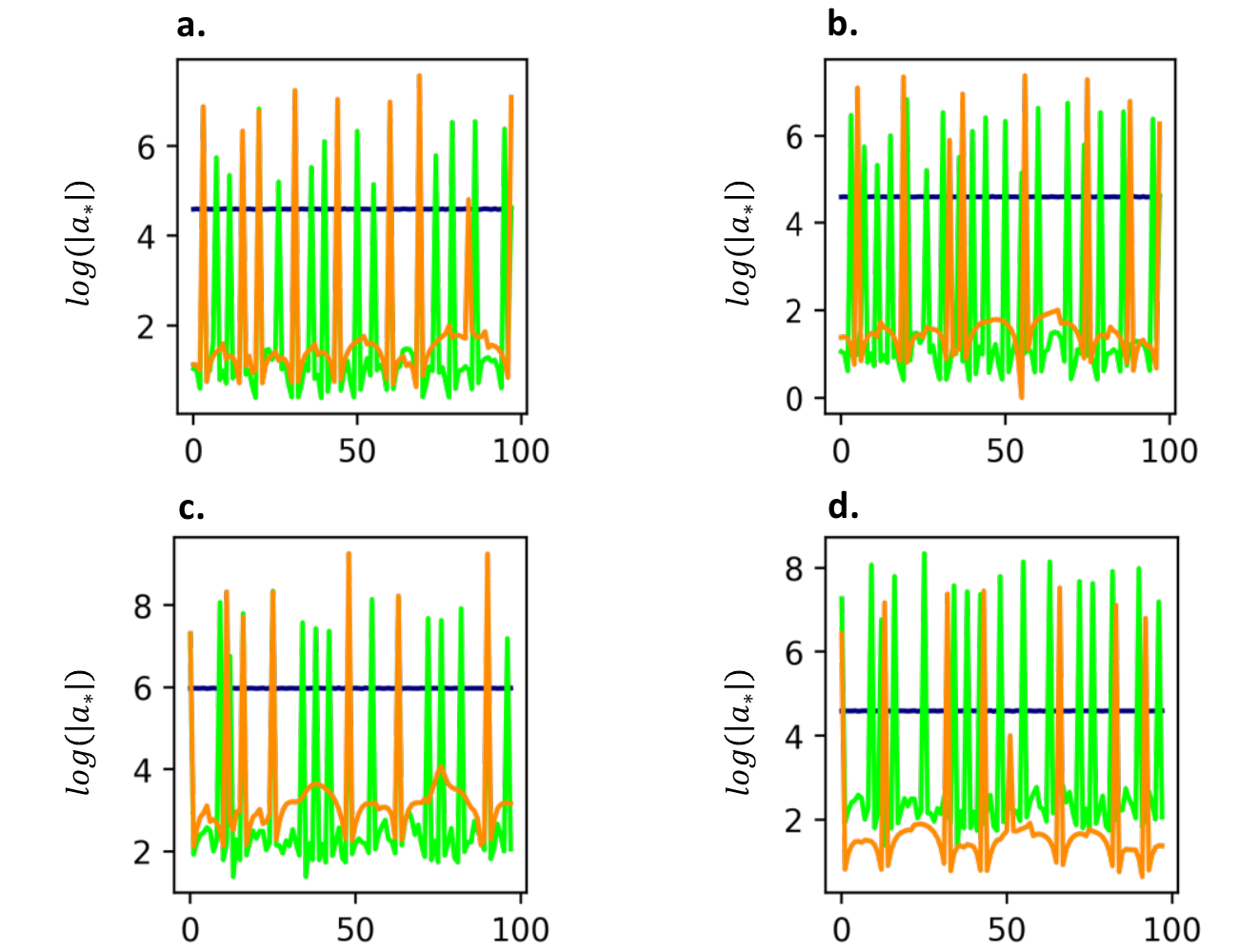}
\caption{\label{fig:eleven} Blue ($t = 0$), green ($t= \text{intermediate}$) and orange ($t=\text{steady state}$) curves represent the value of $a_*$ at the domain boundaries at different time-steps. Panels (a,b): $a_*$ at the top edge for $\mathcal{C_{I_+}} = 100$, $\mathcal{C_{I_-}} = 400$, and $\mathcal{C_{I_+}} = 250$, $\mathcal{C_{I_-}} = 250$, respectively. Panels (c,d): $a_*$ at the bottom edge for $\mathcal{C_{I_+}} = 100$, $\mathcal{C_{I_-}} = 400$, and $\mathcal{C_{I_+}} = 250$, $\mathcal{C_{I_-}} = 250$, respectively.}
\end{figure}

This dependency is further shown in Figure \ref{fig:eleven}, where the value of $a_*$ at top and bottom edges of the square domain at different time-steps are displayed for both cases. The different values of $a_*$ in time at domain boundaries indicate that, the steady-state solutions with the same value of parameters ($K_* = 5\times 10^{-5}$) are different because of the distinct time-varying boundary conditions for $a_*$. Therefore, the model can be simulated using the two fields of supply and drainage density with simple boundary conditions for the densities of the materials. This way, the results can be interpreted in simple terms as the interplay of two indices of the supply and drainage density fields. If the 2-field model is employed, the time-dependent boundary conditions for $a_*$ are extremely complex, and in practice, the obtained co-existing networks can only be constructed from each of the two fields from which the sum $a_*$ originates.
\end{document}